# Mass-energy equivalence for terahertz magnon excitation in antiferromagnetic domain walls

Xu Ge[1], Fa Chen[1], Zaidong Li[2,3], Peng Yan[4*], Hong-Guang Piao[5], Wei Luo[1], Shiheng Liang[6], Xiaofei Yang[1], Long You[1], Yue Zhang[1*]

1. School of Optical and Electronic Information, Huazhong University of Science and Technology, Wuhan, China
2. Department of Applied Physics, Hebei University of Technology, Tianjin, China.
3. Key Laboratory of Electronic Materials and Devices of Tianjin, School of Electronics and Information Engineering, Hebei University of Technology, Tianjin, China
4. School of Electronic Science and Engineering and State Key Laboratory of Electronic Thin Films and Integrated Devices, University of Electronic Science and Technology of China, Chengdu, China
5. College of Science, China Three Gorges University, Yichang, China
6. Department of Physics, Hubei University, Wuhan, China
*Corresponding author: yue-zhang@hust.edu.cn (Yue Zhang); yan@uestc.edu.cn (Peng Yan)

The theory of special relativity is one of the most significant achievements in modern physics, with several important predictions such as time dilation, size contraction for a moving object and mass-energy equivalence. Recent studies have demonstrated size contraction for an antiferromagnetic (AFM) domain wall (DW). Here, we show the mass-energy equivalence by numerically investigating the excitation of terahertz (THz) magnons from a moving AFM DW under the magnetic anisotropy energy gradient. The energy of magnons comes from the loss of DW mass, accompanied with a DW width broadening, overcoming the Lorentz contraction effect. Our results pave the way to study relativistic physics in AFM textures and to efficiently generate THz magnons by electric means.

The theory of special relativity successfully depicts the motion of particles with a velocity that is close to the speed of light. Its application for massive elementary particles gives rise to revolutionary techniques like nuclear plants and weapons. In addition to massive particles, some quasi-particles also obey a relativistic physical principle. A typical example is a terahertz (THz) antiferromagnetic (AFM) magnon/spin wave, which satisfies relativistic dispersion relationship due to the Lorentz invariance of the AFM Lagrangian [1] and has potential application in information technology and biological medicine [1–5].

In general, a spin wave can be excited under external driving force like alternating magnetic field. However, a moving magnetic domain wall (DW) is also a promising candidate for emitting spin waves. The spin-wave emission from a moving ferromagnetic (FM) DW has been intensively investigated [6–8]. A fast FM DW motion can generate spin waves by Cherenkov radiation, but limited by Walker breakdown [7, 8]. Nevertheless, the investigation of AFM DW propagation induced spin-wave emission is still rare, partly because of the lack of an easy driving and detecting method [1]. Thanks to the technique progress in spintronics and multiferroics, the motion of a collinear AFM DW triggered by spin-orbit torque (SOT), electric field, spin wave, or temperature gradient has recently been theoretically investigated [9–18] and experimentally observed [19, 20].

A moving AFM (or ferrimagnetic) DW also behaves as a relativistic massive particle: the DW velocity may not exceed the maximum group velocity of the AFM spin wave ($c_s$) [11, 21, 22]. When the DW velocity approaches $c_s$, the DW experiences significant size shrinking (Lorentz contraction), which boosts the THz spin-wave emission [11, 21]. Very recently, Carreta et al experimentally confirmed the relativistic dynamics of an AFM DW [22].

Besides Lorentz contraction, the mass-energy equivalence is another important prediction of special relativity. An AFM DW is analogous to a relativistic massive object with static self-energy expressed as $(E_{\mathrm{DW}})_0 = m_0 c_s^2$ with $m_0 = \dfrac{2 S_\perp \sqrt{AK}}{c_s^2}$ the static DW mass. Here $A$, $K$, and $S_\perp$ are the homogeneous exchange energy constant, anisotropy constant, and the cross-sectional area of an AFM chain [23], respectively. In this Letter, considering the relativistic behavior of the AFM DW, we propose the self-energy of a moving AFM DW ($E_{\mathrm{DW}}$) as (see the detailed derivation in S1 of the Supplementary Materials):

$$E_{\mathrm{DW}} = \frac{m_0 c_s^2}{\sqrt{1 - \dfrac{v^2}{c_s^2}}}. \tag{1}$$

In nuclear physics, Eq. (1) has been successfully exploited to convert the *real* mass of an object into energy in the form of photons, neutrinos, etc. We thus envision that, in an AFM texture, the loss of the DW *effective* mass can be converted into energy carried by magnons.

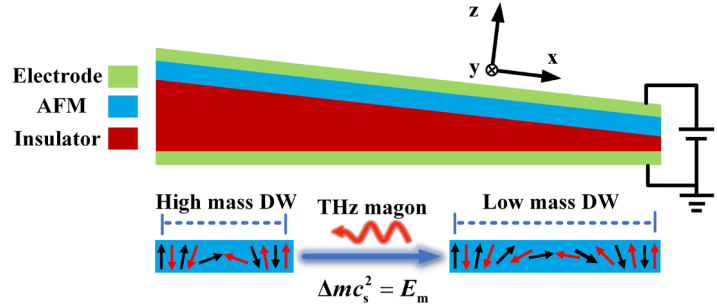

**Figure 1. Schematic of theoretical model: Excitation of THz magnon from a moving DW triggered by voltage-controlled magnetic anisotropy gradient. The AFM DW loses the mass with the reducing anisotropy energy, and the released DW self-energy is converted into THz magnon.**

By numerically studying the dynamics of an AFM DW in the presence of the magnetic anisotropy gradient, we predict a relativistic mass-energy equivalence for THz magnon excitation associated with DW mass reduction. This magnon exhibits discrete spectrum, which is very different from conventional AFM spin wave.

As the mass of an AFM DW is proportional to the square root of anisotropy constant [23], the DW loses its mass with the reducing anisotropy energy. This can be realized via the technique as displayed in Fig. 1: A DC voltage applied across the wedge-shaped insulating layer generates the magnetic energy gradient, which triggers the DW to move towards the end with a lower magnetic anisotropy constant. This method for manipulating anisotropy is not only put forward in theory [24–27] but also realized in experiment [28].

The DW is generated in a 1D AFM chain [dimension: 252 nm (length) × 0.42 nm (width) × 0.42 nm (thickness)] that consists of two FM sublattices. The Hamiltonian of the 1D AFM chain is

$$H = J\sum_{i} \vec{m}^{(i)} \cdot \vec{m}^{(i+1)} - Kd^3 \sum_{i}(\vec{m}^{(i)} \cdot \vec{e}_z)^2 - K_{\perp}d^3 \sum_{i}(\vec{m}^{(i)} \cdot \vec{e}_x)^2. \tag{2}$$

Here $\vec{m}^{(i)}$ is the normalized magnetization at mesh $i$. The first term on the right-hand side of Eq. (2) is the exchange energy with an exchange integral $J$ ($J > 0$). The second term represents the easy-axis anisotropy energy with an anisotropy constant $K$ ($K > 0$). We also consider the hard-axis anisotropy energy in the $yz$ plane with a coefficient $K_{\perp}$ ($K_{\perp} > 0$) due to the shape anisotropy. This hard-axis anisotropy energy is equal to effective easy-axis anisotropy energy along the $x$ direction and gives birth to a Néel-type AFM DW as the initial equilibrium state. In numerical calculations, we choose $d = 0.42$ nm, the lattice constant of NiO [10, 29].

The dynamics of $\vec{m}^{(i)}$ at mesh $i$ is governed by the Landau-Lifshitz-Gilbert equation:

$$\frac{\partial \vec{m}^{(i)}}{\partial t} = -\gamma \vec{m}^{(i)} \times \vec{H}_{eff}^{(i)} + \alpha \vec{m}^{(i)} \times \frac{\partial \vec{m}^{(i)}}{\partial t}, \tag{3}$$

where $\gamma$, $\alpha$, and $\vec{H}_{eff}^{(i)}$ denote the gyromagnetic ratio of an electron, Gilbert damping parameter, and effective field ($\vec{H}_{eff}^{(i)} = -\frac{\delta H}{\mu_0 M_S d^3 \delta \vec{m}^{(i)}}$), respectively. Here $\mu_0$ and $M_S$ are the permeability of vacuum and saturation magnetization. Equation (3) is numerically solved using the Runge–Kutta–Fehlberg method with an adaptive time step (the minimum and maximum time steps are 1 fs and 0.5 ps, respectively) to ensure a small deviation of $m^2$ from 1 ($< 10^{-10}$). We exploit the magnetic parameters of NiO with $M_S = 4.25 \times 10^5$ A/m (1.7 $\mu_B$ magnetic moment for each Ni ion, two spin-up Ni ions, and two spin-down ions in a mesh). $J = 2.1 \times 10^{-22}$ J [29–32], $K = K_0 + s(x - 21.84$ nm), where $K_0 = 3.8 \times 10^5$ J/m$^3$ is the anisotropy constant at $x = 21.84$ nm, the initial position of DW. $s$ is the slope of anisotropy constant, ranging from $-200$ to $-1400$ GJ/m$^4$. As estimated from the shape-anisotropy energy, $K_{\perp}$ is approximately $5.67 \times 10^4$ J/m$^3$. To verify the mass-energy equivalence, we neglect the damping in the main text, but consider its influence in the Supplementary Materials (S1 in the Supplementary Materials).

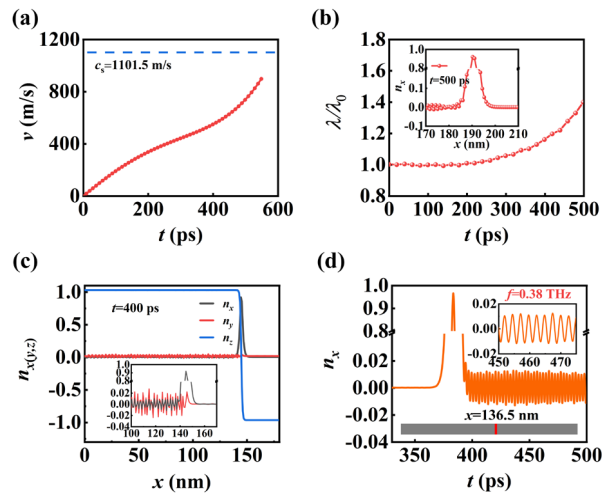

**Figure 2** (a) Temporal DW velocity at $s = -1400$ GJ/m$^4$. (b) Variation of DW width at $s = -1400$ GJ/m$^4$. (c) Spatial $n_x$, $n_y$, and $n_z$ for $s = -1400$ GJ/m$^4$ at 400 ps (Inset: enlarged figure for $n_x$ and $n_y$). (d) Temporal $n_x$ at $x = 136.5$ nm (Inset: enlarged figure for the oscillation of $n_x$ in a short time range with frequency being 0.38 THz.)

We depict the AFM dynamics using the Néel vector $\vec{n}(\vec{r},t) = M_S[\vec{m}_I(\vec{r},t) - \vec{m}_{II}(\vec{r},t)]/l(\vec{r},t)$. Here, $l(\vec{r},t) = M_S|\vec{m}_I(\vec{r},t) - \vec{m}_{II}(\vec{r},t)| \approx 2M_S$.

We simulated the DW motion under different $s$ and found magnon excitation when $s$ is −400 GJ/m$^4$ or greater. Representative magnon excitation from a moving AFM DW is exhibited in Fig. 2. At a large $s$ (−1400 GJ/m$^4$), the DW velocity continues to increase and approaches 1000 m/s at 500 ps [Fig. 2(a)]. It is noticed that the DW velocity is below $c_s$ (1101.5 m/s. See S1 in the Supplementary Materials for more details about the calculation of $c_s$.). Interestingly, the DW width keeps as a constant in the initial 100 ps and substantially increases afterwards [Fig. 2(b)]. This DW broadening indicates the loss of mass, as the DW width and mass are inversely proportional and proportional to the square root of $K$, respectively [23]. As shown in Fig. 2(c), magnon is excited to the stern of the DW [Fig. 2(c)]. This wake-flow magnon is also seen from the temporal evolution of $n_x$ [Fig. 2(d)]. Before the DW reaches the position $x$ = 136.5 nm labeled by the red bar, the $n_x$ oscillation is negligible. After passing it, the DW excites strong $n_x$ oscillation with a frequency around 0.38 THz [the inset of Fig. 2(d)].

As a comparison, we noted that a wake-type spin wave is also emitted from a moving AFM DW induced by SOT [11, 21]. Under SOT, the DW width significantly shrinks, and the DW shape is asymmetric. The magnon is excited due to enhanced exchange energy on the steeper side of DW profile [11, 21]. In this work, however, the DW expands, in sharp contrast to the SOT-driven case [Fig. 2(b)], and the DW shape keeps almost symmetric due to small variation of $K$ in the DW region [Inset of Fig. 2(b)]. Because of the absence of external force, the magnon excitation towards the DW stern works as a recoil force to accelerate the DW.

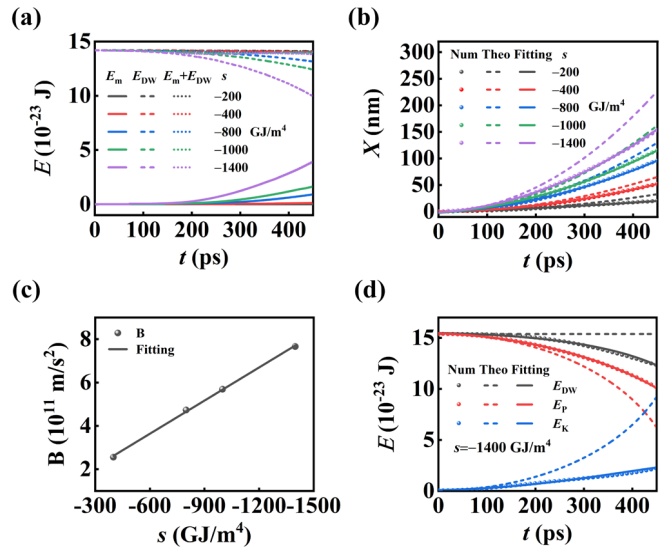

Figure 3. (a) Numerical magnon energy ($E_m$, the solid lines), DW energy ($E_{DW}$, the dashed lines), and $E_m + E_{DW}$ (the dotted lines) at different $s$. (b) Temporal DW displacement ($X$) at different $s$ (the dotted, slashed, and solid lines are the numerical, theoretical solution [Eq. (4)], and the fitting results using Eq. (6), respectively.). (c) Coefficient $B$ in Eq. (6) as a function of $s$. (d) Numerical energy, theoretical energy from Eq. (5), and the fitting energy from Eq. (6) (The black, red, and blue curves are the total DW energy, the self-energy for a static DW, and the kinetic energy of the DW, respectively.)

Figure 3 shows the calculated $E_{DW}$ and the magnon energy ($E_m$). The magnon spectrum will be analyzed below (see Fig. 4). When $s$ is small, $E_m$ is negligible. When $s$ is larger than $-400$ GJ/m$^4$, the magnon starts to emit, which is accompanied by the reducing of $E_{DW}$. It is noteworthy that $E_m$ plus $E_{DW}$ is a constant that equals to the initial DW self-energy, independent on $s$. This indicates that the magnon excitation originates from the release of DW self-energy and satisfies Einstein's mass-energy equivalence.

In what follows, we analytically analyzed the entanglement of magnon excitation and DW propagation. In general, the DW motion is depicted by the dynamics equations of collective coordinates including the DW central position ($X$), the azimuthal angle ($\phi$) of the Néel vector in the DW central, and the DW width ($\lambda$) [33]. A popular assumption in the literature is that $X$ is independent on $\phi$ [1, 9 – 11]. Under this assumption, the DW translation is analogous to the motion of a massive relativistic object obeying the Lagrangian ($L$): $L = -m_0 c_s^2 \sqrt{1 - \frac{\dot{X}^2}{c_s^2}}$ (S3 in the Supplementary Materials) [33]. Based on the procedure in the Supplementary Materials (S3), we derived the temporal $X$ and $E_{DW}$ as:

$$X(t) \approx -\frac{s c_s^2}{4 K_0} t^2, \qquad (4)$$

and

$$E_{DW}(t) \approx 2 S_\perp c_s \sqrt{A} \frac{\sqrt{K_0 + s X(t)}}{\sqrt{c_s^2 + [s c_s^2 X(t) / K_0]}} = m_0 c_s^2. \qquad (5)$$

Here the initial DW velocity is neglected.

As shown in Fig. 3(b), the theoretically predicted DW motion [Eq. (4)] is faster than that of the numerical calculation, especially at a large $s$. More importantly, the $E_{DW}$ calculated by Eq. (5) does not evolve with time [the black slashed line in Fig. 3(d)]. This indicates the released DW self-energy will be totally converted to the DW kinetic energy [the red and blue slashed lines in Fig. 3(d)]. Under this circumstance, the energy-conservation law forbids the magnon excitation.

The inconsistency between Eq. (5) and the numerical calculation comes from the oversimplified assumption of the irrelevance between $X$ and $\phi$. To correct it, we introduce a coupling term $s\dot{X}\dot{\phi}$ and derive the dynamics equation of $X$ and $\phi$ (S4 in the Supplementary Materials). It is straightforward to note that this coupling term vanishes at zero anisotropy gradient. Interestingly, the coupling between DW translation and DW precession indicates that the $E_{DW}$ cannot be fully converted into the DW kinetic energy, leading to the slower numerical DW motion as compared to that calculated by Equation (4). The rest of the $E_{DW}$ is transferred to DW precession and excitation of magnons.

A rigorous solution of the coupled dynamics equation is highly unlikely (Eqs. S18 ~ S20 in S4 in the Supplementary Materials). Nevertheless, the influence of magnon excitation on DW translation can be still estimated as an effective viscosity force that slows down the DW motion. To this end, we assume the DW displacement still obeys a quadratic function of time and propose a fitting formula for $X(t)$ as:

$$X(t) = Bt^2 \qquad (6).$$

It can be seen that the numerical $X(t)$ can be well fitted using Equation (6) [Figure 3(b)] with a coefficient $B$ satisfying a linear relationship to $s$ as: $dB/ds = -0.51$ m$^5$/s$^2$J [Figure 3(c)].

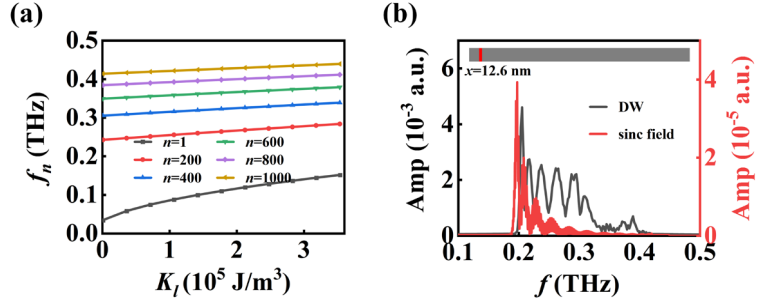

Figure 4. (a) Frequency at the n$^{th}$ level $f_n$ as a function of $K_l$ in an AFM chain with anisotropy gradient. (b) Frequency spectrum for the magnon excited by an AC field of a sinc function (the red line) and by a moving AFM DW (the black line).

Unlike a classical plane spin wave with a continuous spectrum, the anisotropy gradient gives rise to a discrete frequency spectrum:

$$f_n = \frac{c_s}{2\pi}\sqrt{\frac{K_l}{A} + \left(\frac{|s|}{A}\right)^{2/3}\left[\frac{3\pi}{4}(2n-1)\right]^{2/3}}, \quad n = 1, 2, 3 \ldots. \qquad (7).$$

(S5 in the Supplementary Materials). Here $K_l$ is the minimum anisotropy constant in the sloped $K$. The $f_n$ as a function of $K_l$ [Fig. 4(a)] indicates the ground-state frequency $f_1$ is around 0.1 THz, and more than 1000 levels are distributed between 0.1 and 0.4 THz. This means very narrow frequency interval between the neighboring levels.

Importantly, we note that the AFM DW motion can generate magnons with a very broad band. This can be clearly observed in Fig. 4(b) (black curve) in which the magnon amplitude was characterized via the 1-dimentional Fast Fourier Transformation (FFT) for the $n_x$ oscillation at a position ($x = 12.6$ nm) near the left end of the chain. As a comparison, we use a different approach to excite magnons by an AC magnetic field of sinc function (S6 in the Supplementary Materials). We then find that the leading magnon modes appear only in a narrow window 0.2-0.25 THz [see red curve in Fig. 4(b)]. This result shows that the AFM DW propagation offers a potential source to generate broadband THz magnons.

It is well known that atomic energy is huge because of the high light speed. Similarly, an AFM DW with a high $c_s$ also contains significant self-energy [Fig. S1(a) in the Supplementary Materials (S2)]. On the other hand, the electric-field strength for generating the anisotropy gradient can be at the magnitude of $10^{-3}$ V/nm (S7 in the Supplementary Materials), and the voltage is around 1 V for the piezoelectric insulating layer with a thickness of about 1 $\mu$m. This shows that the THz magnon can be excited under a moderate voltage. Finally, we demonstrate the inertial motion of the AFM DW when it leaves the anisotropy gradient region and enters a constant anisotropy region or when it hits the end of the nanowire (S8 in the Supplementary Materials). This needs to be taken into account in real-world applications.

In summary, we predicted the excitation of THz magnons by a moving AFM DW under the magnetic anisotropy gradient. The magnon excitation originates from the dynamic coupling of DW propagation and precession which was largely omitted by the community. We prove that the energy transfer between magnon and DW satisfies Einstein's mass-energy equivalence. Our work paves the way to study relativistic effect based on AFM textures and opens the door for broadband THz magnon generation.

**Acknowledgment**

The authors acknowledge financial support from the National Natural Science Foundation of China (Nos. 51971098, 11874169, and 12074057).

**Reference**

[1] V. Baltz, A. Manchon, M. Tsoi, T. Moriyama, T. Ono, and Y. Tserkovnyak, Rev. Mod. Phys. **90**, 015005 (2018).

[2] B. F. Ferguson and X.-C. Zhang, Nat. Mater. **1**, 26 (2002).

[3] J. Walowski and M. Münzenberg, J. Appl. Phys. **120**, 140901 (2016).

[4] Z. Feng, H. Qiu, D. Wang, C. Zhang, S. Sun, B. Jin, and W. Tan, J. Appl. Phys. **129**, 010901 (2021).

[5] A. Kirilyuk, A. V. Kimel, and T. Rasing, Rev. Mod. Phys. **82**, 2731 (2010).

[6] X. S. Wang, P. Yan, Y. H. Shen, G. E. Bauer, and X. R. Wang, Phys. Rev. Lett. **109**, 167209 (2012).

[7] M. Yan, C. Andreas, A. Kákay, F. García-Sánchez, and R. Hertel, Appl. Phys. Lett. **99**, 122505 (2011).

[8] X. P. Ma, J. Zheng, H. G. Piao, D. H. Kim, and P. Fisher, Appl. Phys. Lett. **117**, 062402 (2020).

[9] W. Yu, J. Lan, and J. Xiao, Phys. Rev. B **98**, 144422 (2018).

[10] E. G. Tveten, A. Qaiumzadeh, and A. Brataas, Phys. Rev. Lett. **112**, 147204 (2014).

[11] T. Shiino, S. H. Oh, P. M. Haney, S. W. Lee, G. Go, B. G. Park, and K. J. Lee, Phys. Rev. Lett. **117**, 087203 (2016).

[12] L. Sánchez-Tejerina, V. Puliafito, P. Khalili Amiri, M. Carpentieri, and G. Finocchio, Phys. Rev. B **101**, 014433 (2020).

[13] H. Y. Yuan, W. Wang, M.-H. Yung, and X. R. Wang, Phys. Rev. B **97**, 214434 (2018).

[14] O. Gomonay, T. Jungwirth, and J. Sinova, Phys. Rev. Lett. **117**, 017202 (2016).

[15] H. Yang, H. Y. Yuan, M. Yan, H. W. Zhang, and P. Yan, Phys. Rev. B **100**, 024407 (2019).

[16] D. L. Wen, Z. Y. Chen, W. H. Li, M. H. Qin, D. Y. Chen, Z. Fan, M. Zeng, X. B. Lu, X. S. Gao, and J.-M. Liu, Phys. Rev. Res. **2**, 013166 (2020).

[17] P. Shen, Y. Tserkovnyak, and S. K. Kim, J. Appl. Phys. **127**, 223905 (2020).

[18] R. Yanes, M. R. Rosa, and L. Lopez-Diaz, Phys. Rev. B **102**, 134424 (2020)

[19] I. Gray, T. Moriyama, N. Sivadas, G. M. Stiehl, J. T. Heron, R. Need, B. J. Kirby, D. H. Low, K. C. Nowack, D. G. Schlom, D. C. Ralph, T. Ono, and G. D. Fuchs, Physical Review X **9**, 041016 (2019).

[20] F. Schreiber, L. Baldrati, C. Schmitt, R. Ramos, E. Saitoh, R. Lebrun, and M. Kläui, Appl. Phys. Lett. **117**, 082401 (2020).

[21] S.-H. Oh, S. K. Kim, D.-K. Lee, G. Go, K.-J. Kim, T. Ono, Y. Tserkovnyak, and K.-J. Lee, Phys. Rev. B **96**, 100407(R) (2017)

[22] L. Caretta, S. H. Oh, T. Fakhrul, D. K. Lee, B. H. Lee, S. K.Kim, C. A. Ross, K. J. Lee, and G. S. Beach, Science **370**, 1438 (2020).

[23] S. K. Kim, Y. Tserkovnyak, and O. Tchernyshyov, Phys. Rev. B **90**, 104406 (2014).

[24] Y. Zhang, S. Luo, X. Yang, and C. Yang, Sci. Rep. **7**, 2047 (2017).

[25] K. Yamada, S. Murayama, and Y. Nakatani, Appl. Phys. Lett. **108**, 202405 (2016).

[26] L. Chen, M. Shen, Y. Peng, X. Liu, W. Luo, X. Yang, L. You, Y. Zhang, J. Phys. D Appl. Phys. **52**, 495001 (2019).

[27] W.H. Li, Z. Jin, D.L. Wen, X.M. Zhang, M.H. Qin, J.-M. Liu, Phys. Rev. B **101**, 024414 (2020).

[28] C. Ma, X. Zhang, J. Xia, M. Ezawa, W. Jiang, T. Ono, S.N. Piramanayagam, A. Morisako, Y. Zhou, and X. Liu, Nano Lett. **19**, 353 (2019).

[29] T. Chatterji, G. J. McIntyre, and P.-A. Lindgard, Phys. Rev. B **79**, 172403 (2009).

[30] T. Archer, C. D. Pemmaraju, S. Sanvito, C. Franchini, J. He, A. Filippetti, P. Delugas, D. Puggioni, V. Fiorentini, R. Tiwari, and P. Majumdar, Phys. Rev. B **84**, 115114 (2011).

[31] D. Ködderitzsch, W. Hergert, W.M. Temmerman, Z. Szotek, A. Ernst, and H. Winter, Phys. Rev. B **66**, 064434 (2002).

[32] W.-B. Zhang, Y.-L. Hu, K.-L. Han, and B.-Y. Tang, Phys. Rev. B **74**, 054421 (2006).

[33] G. Tatara, C. A. Akosa, and R. M. Otxoa de Zuazola, Phys. Rev. Res. **2**, 043226 (2020).